\documentclass[a4paper,12pt]{article}
\usepackage{epsfig,amssymb,amsmath,textcomp,caption,verbatim,gensymb,wrapfig,mathtools, tabularx}
\usepackage{epstopdf, ltablex,booktabs}
\usepackage{graphicx}
\usepackage{amsfonts,braket}
\usepackage[top=1.0in, left=1.0in, bottom=1.5in, right=1.0in]{geometry}
\usepackage[font={small}]{caption}
\usepackage{subfig}
\usepackage[colorlinks=true,citecolor=blue]{hyperref}
\usepackage{cleveref}
\usepackage[square,comma,numbers,compress]{natbib}

\numberwithin{equation}{section}

\newcommand{\website}{\href{http://www1.maths.leeds.ac.uk/pure/geometry/SkyrmionVibrations/}{http://www1.maths.leeds.ac.uk/pure/geometry/SkyrmionVibrations/}}

\begin{document}
	
	\thispagestyle{empty}
	\renewcommand{\thefootnote}{\fnsymbol{footnote}}
	\vskip 3em
	\begin{center}
		{{\bf \Large Vibrational modes of Skyrmions}}\\[15mm]	
		{\large Sven Bjarke Gudnason$^1$\footnote{email:
				gudnason@keio.jp}
			and Chris Halcrow$^2$\footnote{email: c.j.halcrow@leeds.ac.uk}} \\[1pt]
		\vskip 1em
		{\it $^1$Department of Physics, and Research and Education Center for Natural Sciences, \\ Keio University, Hiyoshi 4-1-1, Yokohama, Kanagawa 223-8521, Japan,}\\
		{\it $^2$School of Mathematics, University of Leeds, Leeds, LS2 9JT, UK.}
		
		\bigskip
		\bigskip
		{\bf Abstract}
	\end{center}
	We study the vibrational modes of Skyrmions with baryon numbers one
	through eight in the standard Skyrme model. The vibrational modes are
	found in the harmonic approximation around the classical soliton
	solution and the real parts of the frequencies of the modes are
	extracted. We further classify the vibrational modes into
	representations of the symmetries possessed by the Skyrmion solutions.
	We find that there are approximately $7B$ low-lying modes for a
	Skyrmion with baryon number $B$, in addition to an infinite continuum
	of scattering modes. This result suggests that the instanton moduli space, which is $8B$-dimensional, does not accurately describe the deformation space of Skyrmions as previously conjectured.

	\vfill
	\newpage
	\setcounter{page}{1}
	\setcounter{footnote}{0}
	\renewcommand{\thefootnote}{\arabic{footnote}}

	\section{Introduction}
	
	In the Skyrme model \cite{Sky}, nuclei are described as topological solitons known as Skyrmions. The Skyrmions are classical energy minimizers of a nonlinear field theory and are labeled by a topologically conserved integer, $B$, which is identified with the baryon number. In nuclear physics this is called the atomic mass number and is equal to the number of nucleons. 
	
	To make contact with nuclear physics, one must quantize the Skyrmions. Ideally, all possible field configurations should be included in the quantization scheme. Unfortunately  this is too difficult and instead one must truncate the configuration space to make the calculation tractable. This can also be thought of as truncating the degrees of freedom of the Skyrmions themselves. The simplest and most widely applied truncation is to only include the zero modes, those transformations which leave the Skyrmion's energy unchanged: translations, rotations and isorotations. This procedure has some successes, such as the reproduction of the energy spectra and static properties of some small-$B$ nuclei \cite{ANW,MMW}. However, since these degrees of freedom do not allow the Skyrmion to break up, the formalism cannot be used to study several important problems such as nuclear fission, binding energies and scattering processes. A first step in studying these problems is to understand the normal mode spectra of the Skyrmions. One can then understand how the Skyrmions deform and break apart classically. We call the space of deformed Skyrme configurations the deformation space. Including some of these additional degrees of freedom has led to successes in describing the Deuteron \cite{LMS}, Lithium-$7$ \cite{Hal} and Oxygen-$16$ \cite{HKM} nuclei.
	
	Basic questions about the normal mode spectra are still unanswered. For instance, it is unclear how many normal modes each Skyrmion has. Various authors have theorized that a baryon number $B$ Skyrmion has $6B+1$ \cite{BBT2}, $8B-4$ \cite{BM} and $8B-1$ \cite{AM} modes (including the nine zero modes) based on numerical results, the rational map approximation and instanton ideas respectively. In this paper, we  answer this question by numerically generating the normal modes of the $B=1-8$ Skyrmions. We find that the question is ill-posed and that low lying quasi-normal modes should also be included in the counting. Once these are accounted for, we find that there are approximately $7B$ modes, rejecting all previous conjectures. The result can be understood as follows: as well as its six zero modes, translations and rotations, a single Skyrmion can also increase and decrease its size radially. This is known as the breathing mode. The $7B$-dimensional deformation space of a $B$-Skyrmion includes enough degrees of freedom for its constituent Skyrmions to take advantage of all their seven modes.
	
	The paper is organized as follows. We carefully describe the spectral problem and how we solve it in Section \ref{sec:2}. The overall results and a list of modes can be found in Section \ref{sec:3}, where we also discuss several interesting individual modes. We have generated 266 normal and quasi-normal modes and each may be important for certain nuclear processes. Section \ref{sec:4} contains our conclusions. To keep the paper readable, we have relegated technical details about the symmetries of the modes to Appendix \ref{app:A}. A full list of the 266 modes, accompanied by descriptions, can be found in Appendix \ref{app:B}. These are also available online accompanied by animations of each mode at \website.
	
	\section{Problem and Method}\label{sec:2}
	
	The Skyrme Lagrangian is usually written in terms of an $SU(2)$-valued field, $U(\boldsymbol{x},t)$. For the numerical setup it is more convenient to write the field in terms of a $4$-component nonlinear sigma-model field $\Phi = (\Phi_0, \boldsymbol{\Phi})$ such that
	\begin{equation}
	U = \Phi_0 + i \boldsymbol{\tau}\cdot \boldsymbol{\Phi} \, ,
	\end{equation}
	where $\boldsymbol{\tau}$ are the Pauli matrices and the fields satisfy
	\begin{equation} \label{constraint}
	\Phi_0^2 + \boldsymbol{\Phi}\cdot\boldsymbol{\Phi} = 1 \, ,
	\end{equation}
	so that $U \in SU(2)$. In this notation, the standard Skyrme Lagrangian in natural units is
	\begin{equation} \label{Lag}
	\mathcal{L} = \partial_\mu \Phi \cdot \partial^\mu \Phi + \frac{1}{2}\left( \partial_\mu \Phi \cdot \partial_\nu \Phi \right)\left( \partial^\mu \Phi \cdot \partial^\nu \Phi \right) - \frac{1}{2}\left( \partial_\mu \Phi \cdot \partial^\mu \Phi \right)^2 - m^2 (1-\Phi_0) \, ,
	\end{equation} 
	where $m$ is the dimensionless pion mass which we fix as $m=1$. One may modify the Lagrangian in a number of ways \cite{NS,GHS,Gud,Gud2,Gud3,Gud4,Gud5,ASW}. For static, finite energy configurations the potential term ensures that $\Phi=(1,0,0,0)$ at spatial infinity. This gives a one-point compactification of space and means that $\Phi$ is a map from $\mathbb{R}^3 \cup  \{ \infty \} \cong S^3$ to $SU(2) \cong S^3$. Maps between three-spheres have an associated topological degree, which we denote $B$. The degree is conserved and so Skyrme configurations with different values of $B$ lie in disjoint sectors of the total configuration space. In each sector there is a minimal energy configuration which we call the Skyrmion with charge $B$.
	
\begin{figure}[!ht]
		\begin{center}
			\mbox{\subfloat[$B=1, O(3)$]{\includegraphics[height=0.2\linewidth]{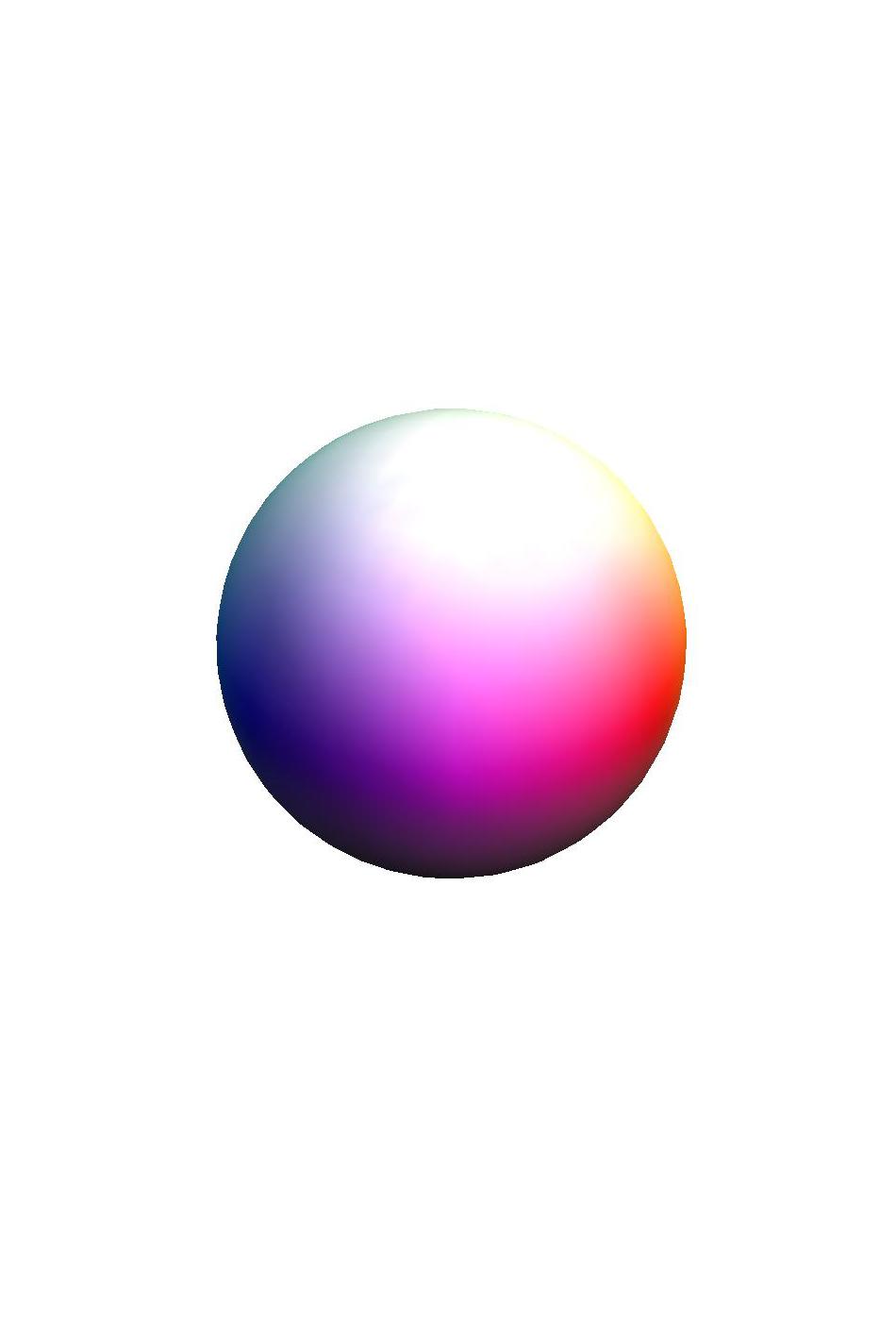}}
				\subfloat[$B=2, D_{\infty h}$]{\includegraphics[height=0.2\linewidth]{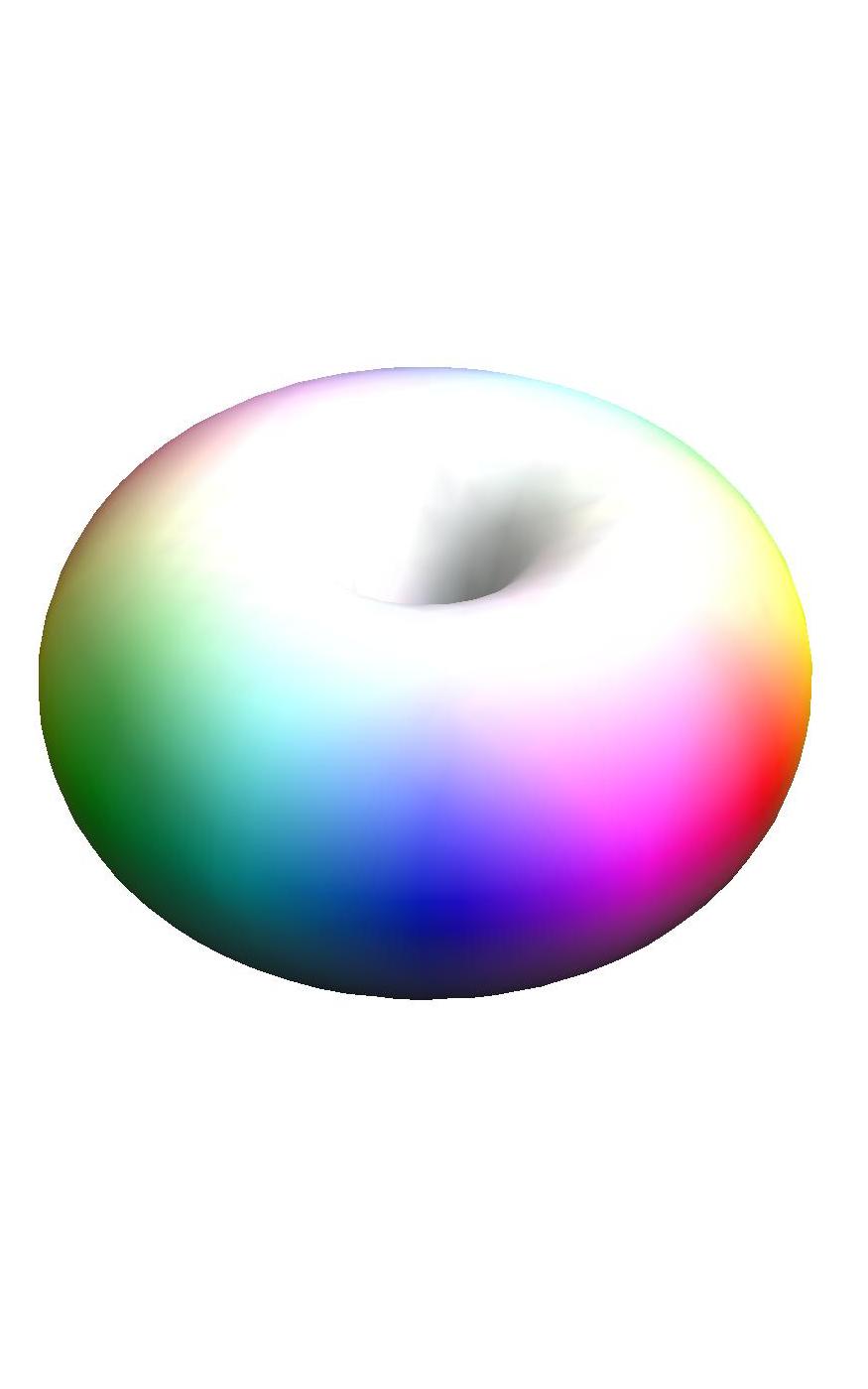}}
				\subfloat[$B=3, T_d$]{\includegraphics[height=0.2\linewidth]{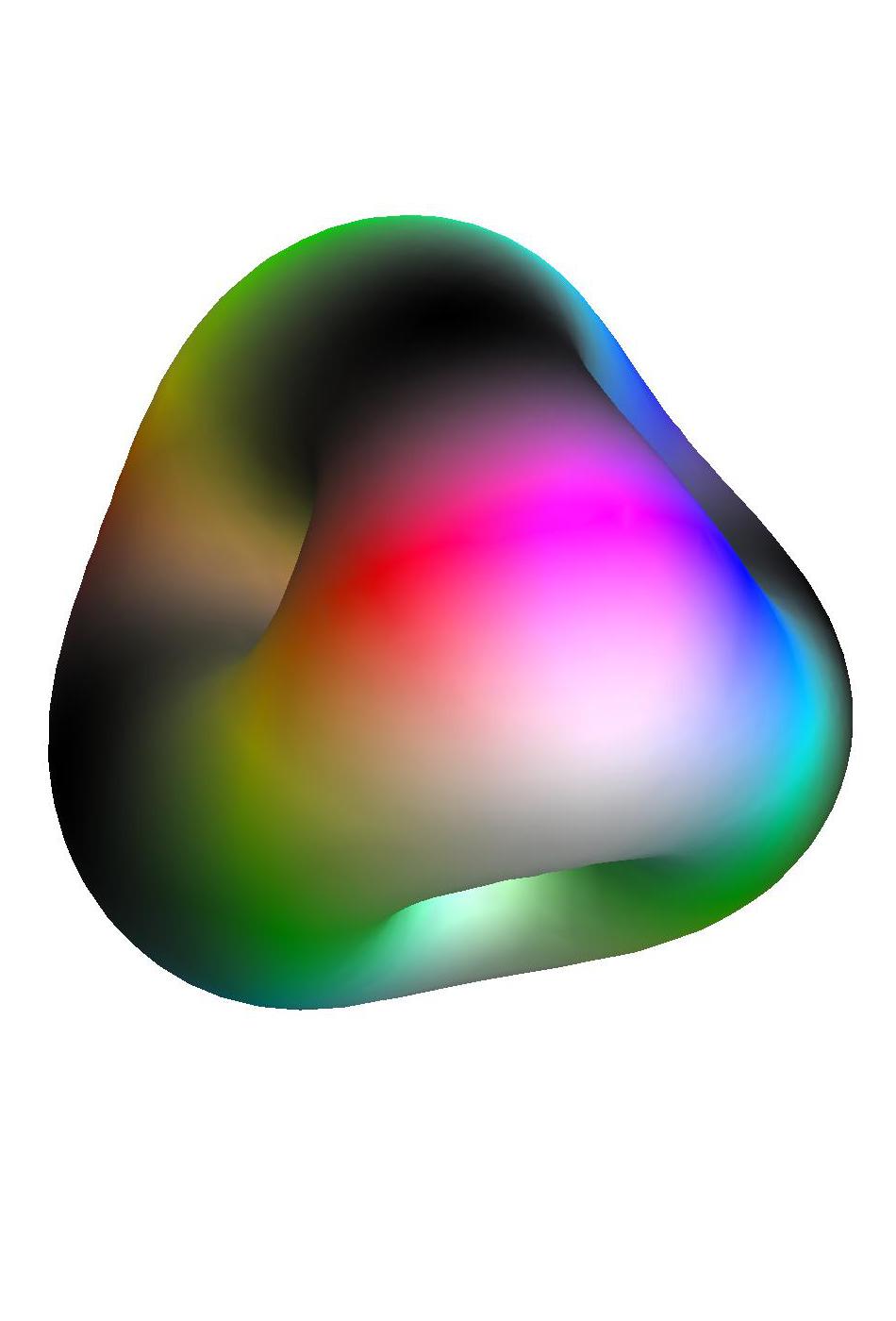}}
				\subfloat[$B=4, O_h$]{\includegraphics[height=0.2\linewidth]{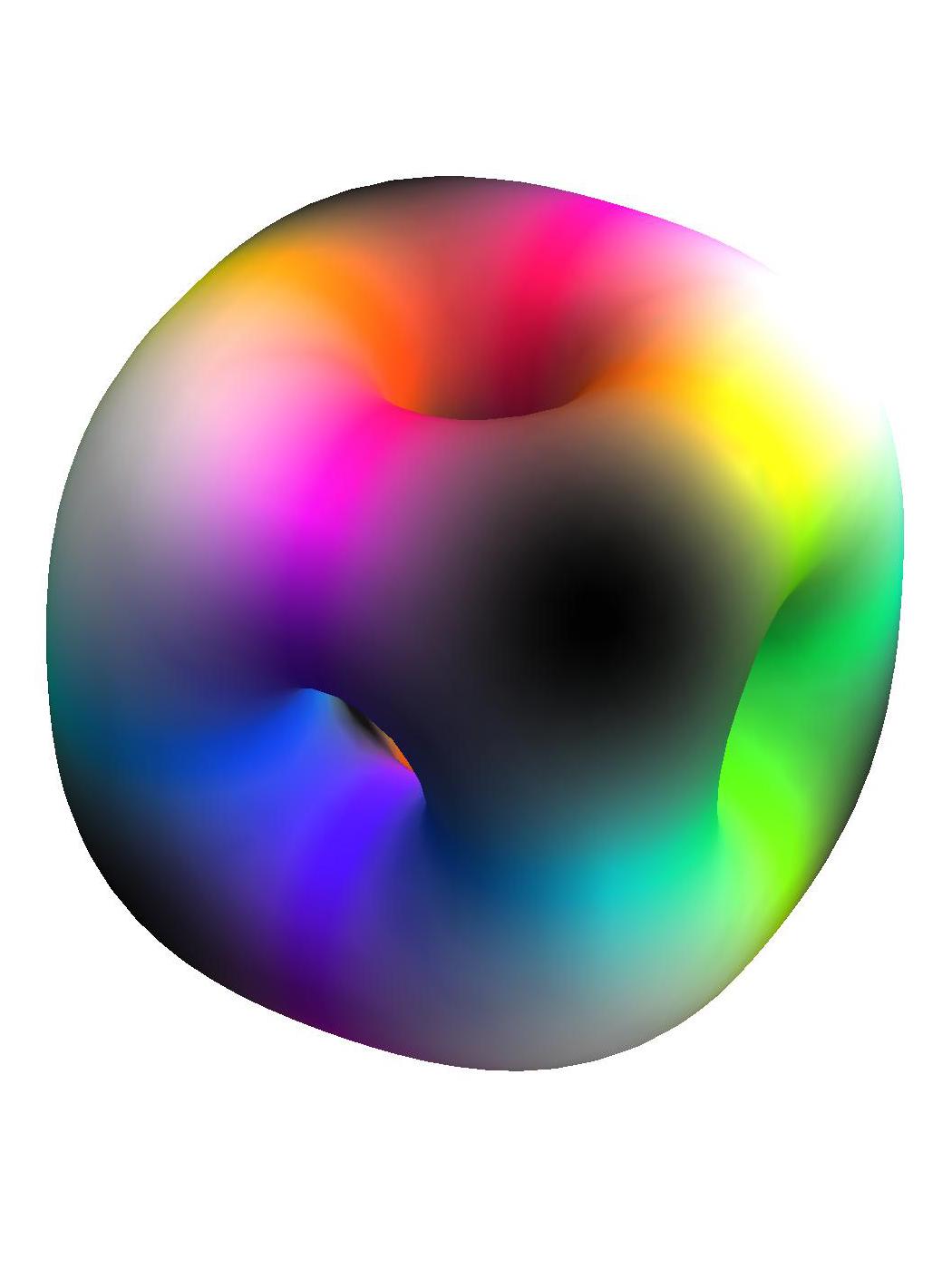}}
				\subfloat[$B=5, D_{2d}$]{\includegraphics[height=0.2\linewidth]{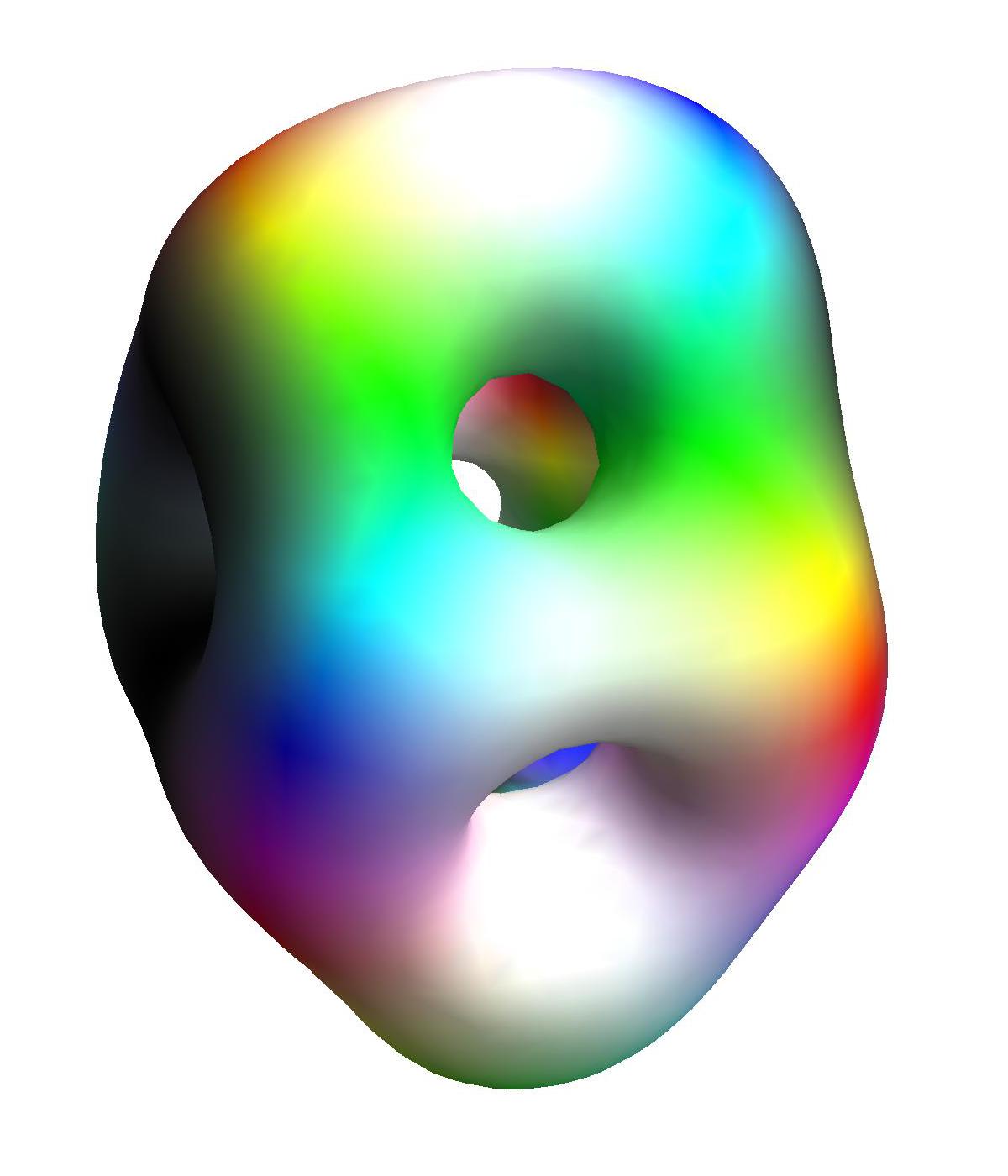}}
				\subfloat[$B=6, D_{4d}$]{\includegraphics[height=0.2\linewidth]{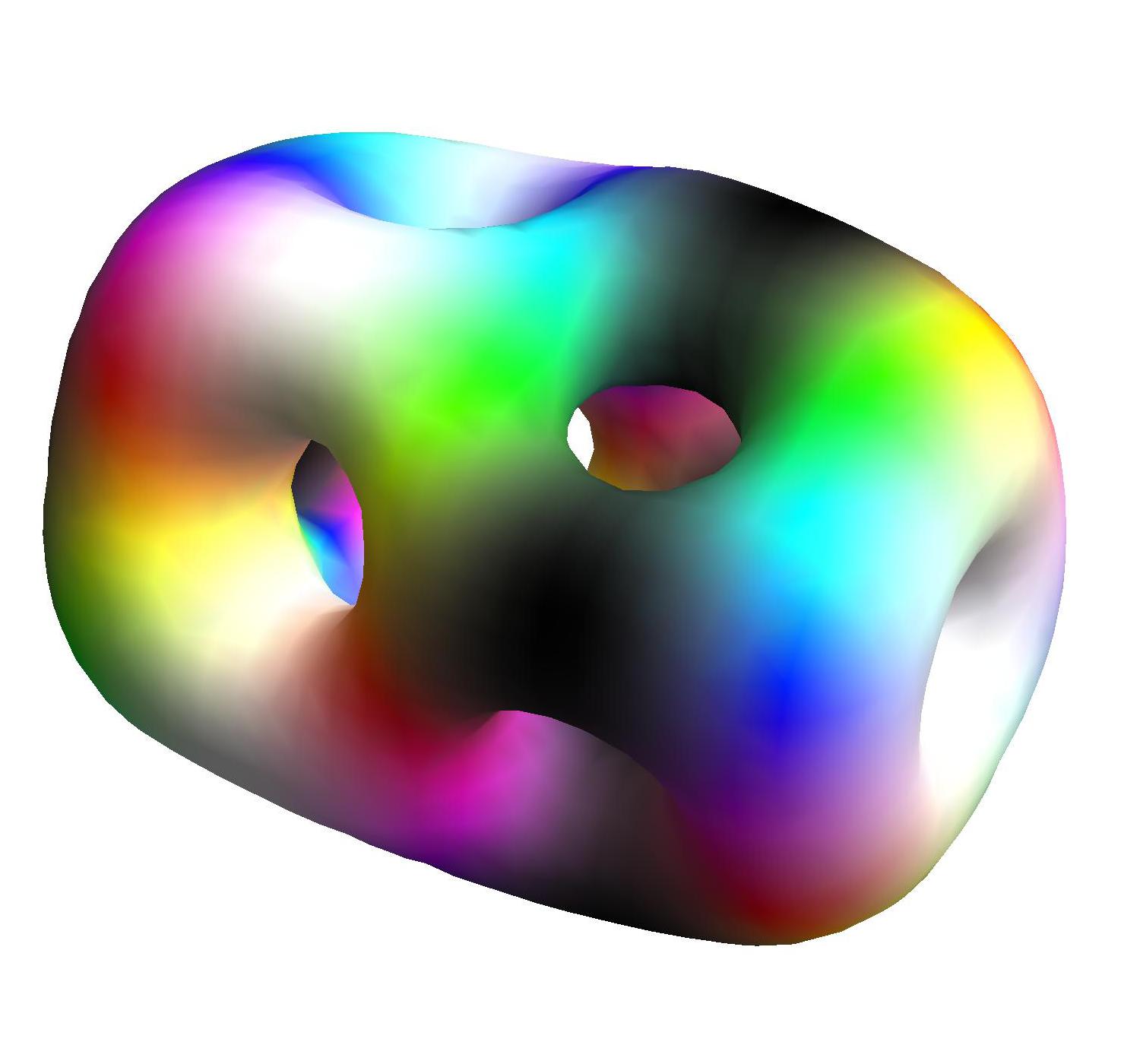}}}
			\mbox{\subfloat[$B=7, I_h$]{\includegraphics[height=0.2\linewidth]{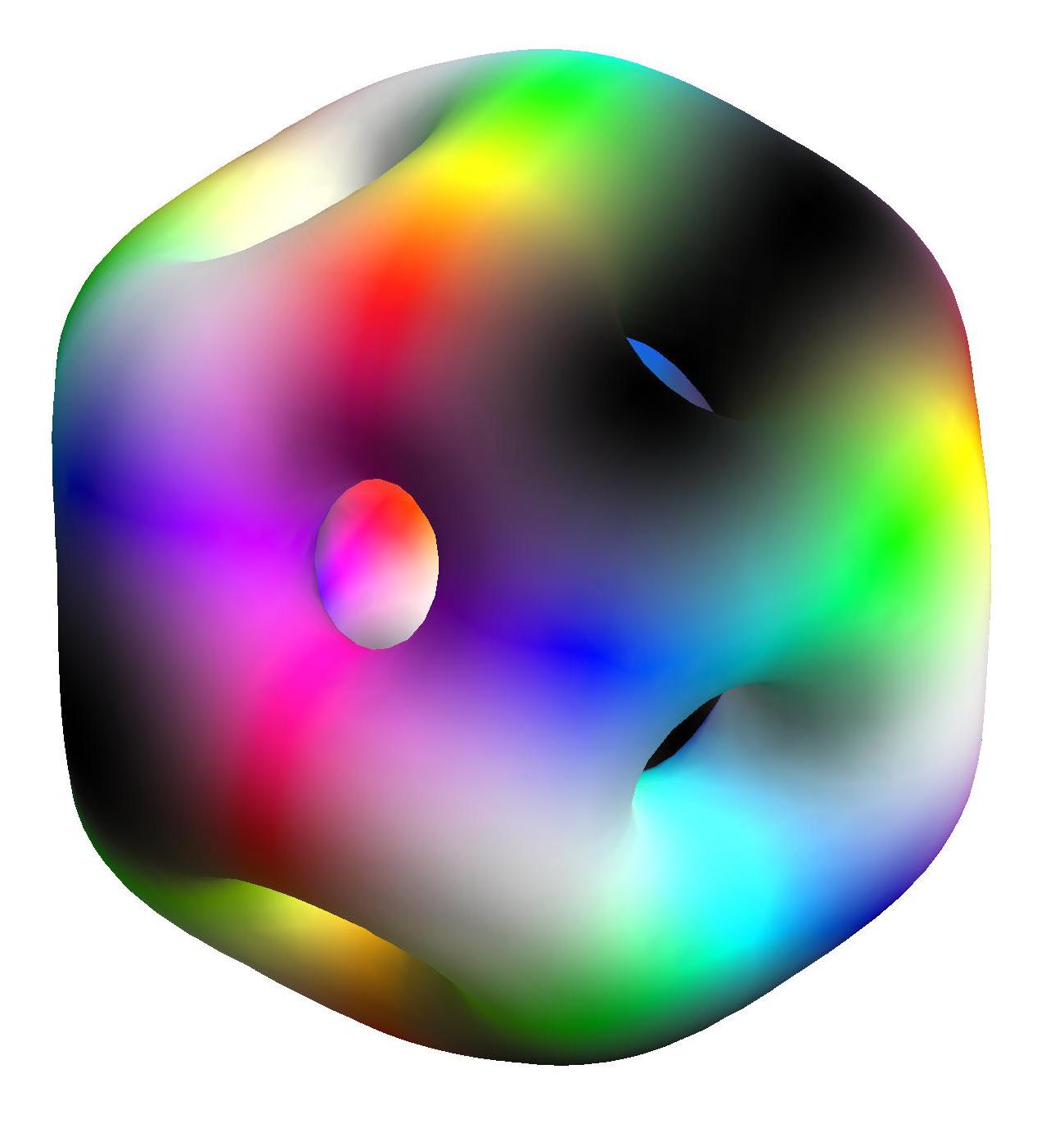}}
				\subfloat[$B=8_h, D_{6d}$]{\includegraphics[height=0.2\linewidth]{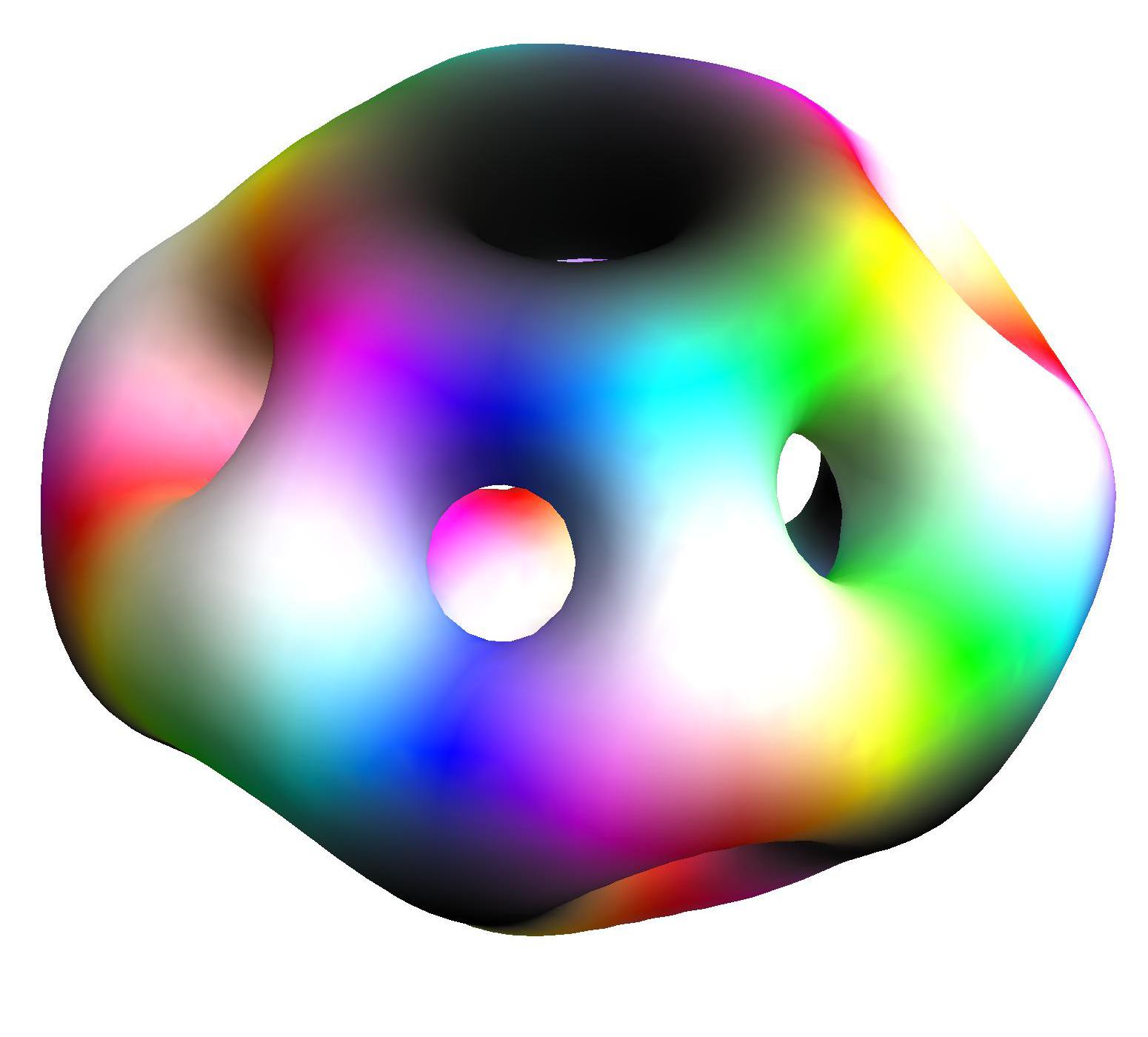}}
				\subfloat[$B=8_t, D_{4h}$]{\includegraphics[height=0.2\linewidth]{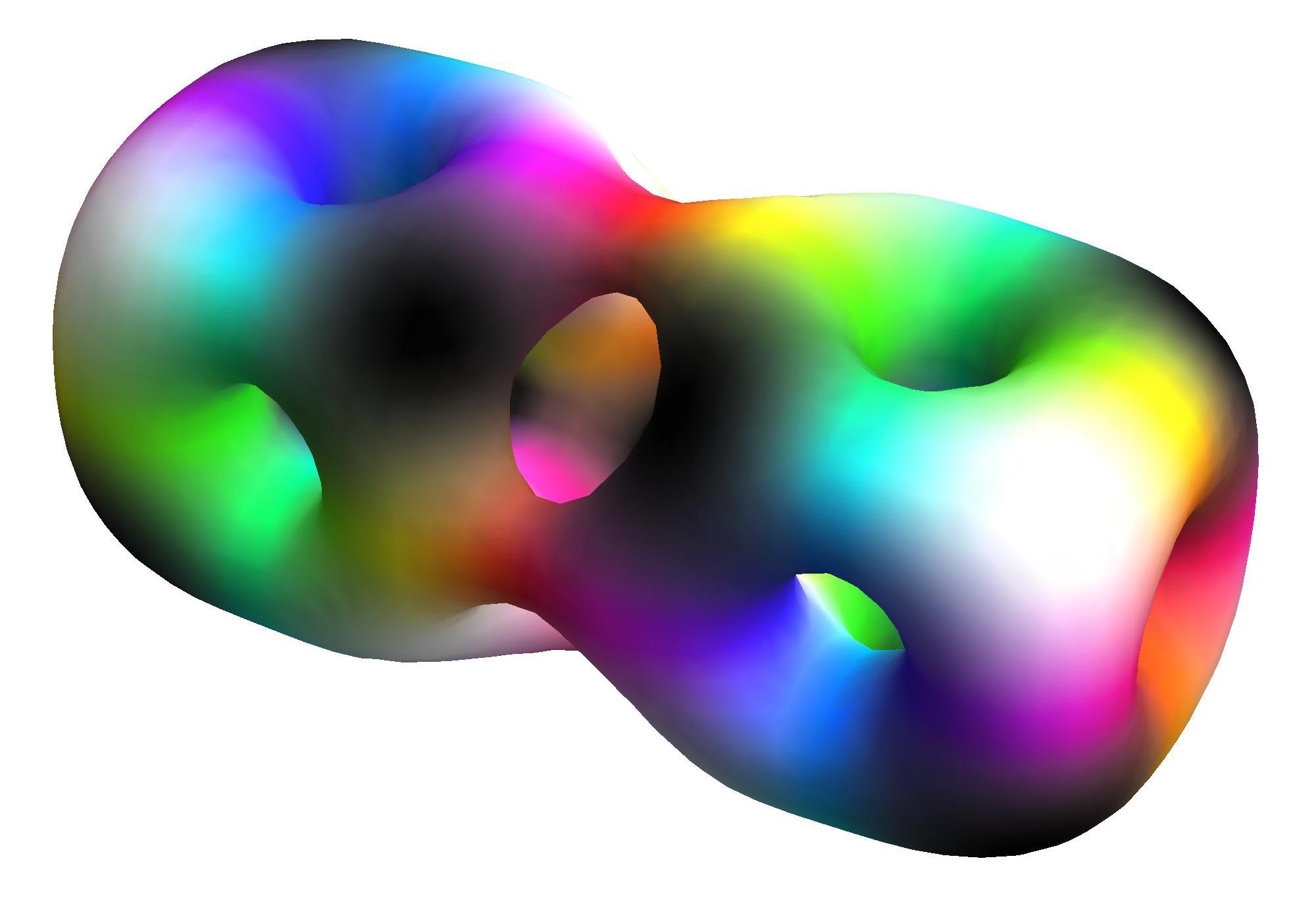}}
				\subfloat[$B=8_u, D_{4h}$]{\includegraphics[height=0.2\linewidth]{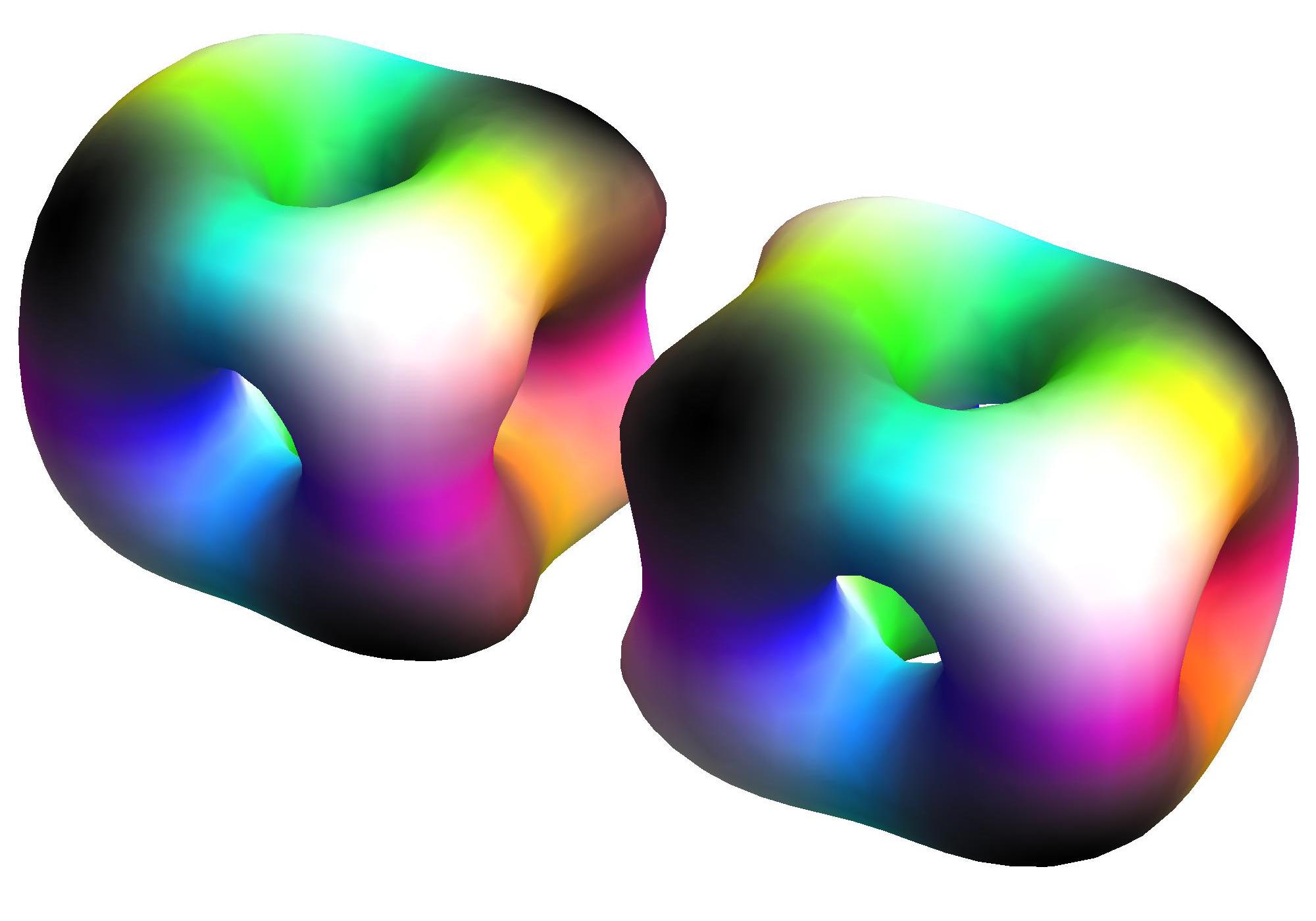}}} 
			\caption{The small-$B$ Skyrmions and their symmetry groups, to scale. }
			\label{fig:Skyrmions}
		\end{center}
	\end{figure}

	To visualize a Skyrme configuration we plot a contour of constant energy density. This is then colored to represent the direction of the pion field at that point on the energy contour. The Skyrme field is colored white/black when $\hat{\pi}_3 = \pm 1$ and red, green and blue when $\hat{\pi}_1+i\hat{\pi}_2 = \exp(0)$, $\exp(2i \pi/3)$ and $\exp(4 i \pi /3)$ respectively, where $\hat{\boldsymbol{\pi}}\equiv\boldsymbol{\pi}/|\boldsymbol{\pi}|$ is the normalized pion field. The small $B$ Skyrmions are well known and have interesting symmetry groups. We plot the $B=1-8$ Skyrmions in Figure \ref{fig:Skyrmions}. Modifications to the Skyrme Lagrangian can alter the symmetries of the Skyrmions. Note that there are three $B=8$ solutions. One has $D_{6d}$ symmetry which we label $B=8_h$. Another is constructed from two $B=4$ Skyrmions which have been rotated relative to each other. This is known as the twisted $B=8$ solution and so we label it $B=8_t$. Finally, there is one saddle point solution constructed from $B=4$ clusters which have not been rotated relative to each other, known as the untwisted solution and labeled $B=8_u$. The three solutions are almost degenerate in their classical energies. 
	
	We denote a Skyrmion solution by $\phi$ and study small deformations of it. To do so, we consider fields of the form
	\begin{equation} \label{modeex}
	\Phi_\epsilon(\boldsymbol{x},t) = \phi(\boldsymbol{x}) + \epsilon(\boldsymbol{x}) e^{i \omega t} \, ,
	\end{equation}
	where $\epsilon(\boldsymbol{x})$ is a normal mode with frequency $\omega$. We take $\epsilon$ to be small and, to ensure that $\Phi_\epsilon$ remains in $SU(2)$, demand that
	\begin{equation} \label{constraint2}
	\phi \cdot \epsilon = 0 \, .
	\end{equation}
	Inserting equation \eqref{modeex} into the full Lagrangian \eqref{Lag} and ignoring higher-order terms in $\epsilon$, we find a linearized Lagrangian for $\epsilon$. The corresponding Euler-Lagrange equation is
	\begin{equation} \label{epeom}
	\partial_i\left(V_{a b; i j} \partial_j \epsilon_b \right) + \lambda \epsilon_a = \omega^2 I_{ab} \epsilon_b \, ,
	\end{equation}
	where
	\begin{align}
	V_{ab;ij} &=  \delta_{ab} \partial_i \phi_c \partial_j \phi_c - \delta_{ij}\delta_{ab} \partial_k \phi_c \partial_k \phi_c + \partial_j \phi_a \partial_i \phi_b+\delta_{ij}\partial_k\phi_a \partial_k\phi_b-2\partial_i \phi_a \partial_j \phi_b - \delta_{ab}\delta_{ij} \, , \nonumber \\
	I_{ab} &= \delta_{ab} + \delta_{ab} \partial_i \phi_c \partial_i \phi_c - \partial_i \phi_a \partial_i \phi_b \, ,
	\end{align}
	and $\lambda$ is the Lagrange multiplier of the solution
	\begin{equation}
	\lambda = \phi \cdot \left( \left( \partial_i \phi \cdot \partial_i\partial_j \phi - \partial_j \phi \cdot \partial_i \partial_i \phi \right)\partial_j \phi + \left(1+\partial_j \phi \cdot \partial_j \phi)\partial_i\partial_i \phi - \left(\partial_i\phi \cdot \partial_j \phi\right) \partial_i \partial_j \phi\right)\right)+m^2 \phi_0 \, ,
	\end{equation}
	which enforces the constraint \eqref{constraint2} in theory. However, due to small numerical errors, the constraint does not hold exactly and these errors are compounded with time. Hence, we manually project the vibration into the plane \eqref{constraint} every few timesteps. Equation \eqref{epeom} is the linearized equation of motion for the normal modes of the classical Skyrmion $\phi$. The equations are in Sturm-Liouville form and thus there exists an eigenfunction basis, also known as the mode space, which we denote $\{ \epsilon^{(p)} \}$. The basis can be made orthogonal with respect to the weight function so that
	\begin{equation}
	\int \epsilon^{(p)}_a I_{ab}(\phi) \epsilon^{(q)}_b\, d^3 \boldsymbol{x} = \delta_{pq}\, .
	\end{equation}
	The local deformation space is generated by all configurations $\Phi = \phi + \sum_n a_n \epsilon^{(n)}$ for real coefficients $\{a_n\}$.
	
	There are three types of mode. Their difference can be understood by examining the equations of motion far from the Skyrmion. Here, the Skyrme field looks like the vacuum state plus a Yukawa multipole and thus equation \eqref{epeom} can be found asymptotically. It becomes
	\begin{equation} \label{asymp}
	\left( -\nabla^2 + m^2\right)\epsilon = \omega^2 \epsilon \, ,
	\end{equation}
	which has solution $\exp\left( \pm \sqrt{m^2-\omega^2} \, \, \boldsymbol{k} \cdot \boldsymbol{x} \right)$ for any constant unit vector $\boldsymbol{k}$. Real bounded solutions only exist for $\omega < m$. Since a normal mode must decay asymptotically they only exist for $\omega<m$. There is then an infinite set of scattering modes with $\omega > m$. Any numerical scheme will naturally discretize this continuous spectrum. In addition, there are quasi-normal modes (QNMs) which have complex frequencies. At a fixed time they are non-normalizable, but they do decay dynamically. QNMs are well understood in black holes, where they are responsible for long-time sphericity of solutions, and quantum mechanics where they describe resonant states. See Ref.~\cite{QNM} for a review or Ref.~\cite{DR} for an example of their existence in a simple soliton model. Since we are in a finite box the spatial blow up does not cause a problem.  Our method only allows us to extract the real part of the frequencies of the QNMs. This paper and its results would be improved by using a method which is tailored to find QNMs. Note that, due to the mass term in equation \eqref{asymp}, a QNM may become a normal mode as $m$ is increased. This has been shown to happen in the case of the $B=1$ Skyrmion \cite{AHRW,AHRW2}. We find repeated evidence that this also occurs for $B>1$. Hence varying $m$ can (and does) alter the number of normal modes which exist. The question ``how many normal modes does a Skyrmion have?" does not have a unique answer. Instead we will ask ``how many low lying normal and quasi-normal modes does a Skyrmion have?" This does appear to have an answer. To do the calculation we must apply a cut-off frequency in our search for modes. We choose $\omega = 1.5m = 1.5$. An obvious extension to this work would be to find the effect of changing this cut-off. We call the combined set of normal modes and QNMs vibrational modes.
	
	From the asymptotic analysis, it appears that no normal modes exist if $m=0$. In fact this is an artificial consequence of linearizing the nonlinear Skyrme theory. The linearization \eqref{epeom} misses the fact that Skyrmions may break up and that the resulting clusters can move arbitrarily far apart. For example, the $B=2$ Skyrmion can break into two $B=1$'s which may be infinitely separated. The equation \eqref{epeom} is focused around the classical toroidal solution and thus cannot describe this mode in full, only those configurations near the torus. If one changed into different coordinates which can accommodate this motion, such as those based on the instanton moduli space \cite{LMS}, the break up mode could be studied in more detail. Whether that break-up corresponds to a normal mode or a QNM will depend on the value of the asymptotic Skyrmion configurations rather than the naive asymptotics given by equation \eqref{asymp}.
	
	To find the vibrational modes we follow a procedure outlined in Ref.~\cite{WW,BT}. We generate an initial random perturbation $\epsilon(\boldsymbol{x})$ at time $\tau = 0$ and evolve it using
	\begin{equation} \label{evolution}
	\partial_\tau\epsilon(\boldsymbol{x},\tau) = -I^{-1}_{ab} \left(\partial_i\left(V_{bc;ij}\partial_j \epsilon_c \right) +\lambda \epsilon_b\right)\, .
	\end{equation}
	This has solution
	\begin{equation}
	\epsilon(\boldsymbol{x},\tau) = \sum_n b_n e^{-\omega_n^2 \tau}\epsilon^{(n)}\, ,
	\end{equation}
	where $\{\epsilon^{(n)} \}$ is the set of normal modes and $b_n$ are some constant coefficients. After a long time, the lowest-frequency mode dominates the solution and we extract it. Having found this mode we can project it out of the initial perturbation (essentially setting $b_1=0$) and repeat the process. Now the solution does not contain any of the lowest frequency mode and after a long time the second-lowest mode dominates. We can then extract this mode and project it (as well as the lowest mode) out of the initial perturbation and repeat the process again. This is repeated until we have found all modes up to the cut off $\omega = 1.5$. To improve the method we take advantage of the symmetries of the Skyrmions. Each mode transforms as an irreducible representation (irrep) of the Skyrmion's symmetry group and we treat each irrep individually by applying symmetry operators to the perturbations. More details, including lists of the irreps, can be found in Appendix \ref{app:A}. We display the dimension of the irrep as a prefixed superscript. Hence the two-dimensional irrep $E$ is written $^2E$.
	
\begin{figure}[!ht]
		\begin{center}
			\includegraphics[width=\linewidth]{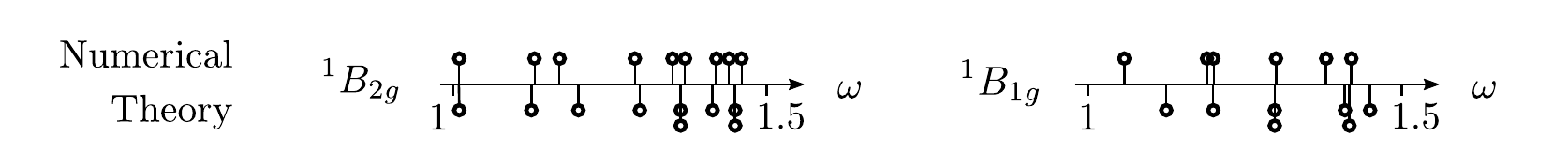}
			\caption{The theoretical spectrum, using the asymptotic scattering states as an approximation (bottom), against the numerically generated spectrum (top) for the $^1B_{2g}$ and $^1B_{1g}$ irreps of the $B=8_t$ Skyrmion. The theory works well for the $^1B_{2g}$ irrep but fails for the $^1B_{1g}$ irrep, providing initial evidence for the existence of a QNM which transforms as $^1B_{1g}$.}
			\label{fig:quasiev}
		\end{center} 
	\end{figure}

	The QNMs pose a greater challenge than the normal modes as they are significantly more difficult to identify and extract. This is because they lie in the same frequency range as, and hence mix with, the scattering modes. We use a variety of techniques to deal with this problem. Each numerical run was repeated on three different box sizes since QNMs are less sensitive to this change compared with scattering states \cite{Mai}. We also estimate the  scattering modes by solutions of the asymptotic equation \eqref{asymp}. This is a surprisingly good approximation when there are no QNMs present in the spectrum but the spectrum is greatly altered when one is present. The theoretical spectrum and the numerically generated spectrum for the $B=8_t$ Skyrmion, for two different irreps, is plotted in Figure \ref{fig:quasiev}. The first irrep, $^1B_{2g}$, does not contain a QNM while the second, $^1B_{1g}$, does. In the Figure, the theoretical scattering modes match the numerical results very well for $^1B_{2g}$ but the matching fails entirely for the $^1B_{1g}$ modes. Spectra like this gives indirect evidence of the existence of a QNM. One can also compare the structure of the approximate scattering modes with the numerical modes to confirm an identification. If these techniques do not give clear results, we then repeat the calculation at a higher value of $m$. For a large enough $m$ the QNM becomes normal and one can then confirm that it is indeed a QNM rather than a scattering mode.
	
	In total, we investigated and classified 655 modes, 266 of which were identified as vibrational modes and 86 as zero modes. All numerical simulations were done on a cuboid with periodic boundary conditions using fourth-order spatial derivatives and a lattice spacing of $0.15$ Skyrme units.  
	
	\section{Results} \label{sec:3}
	
	We now summarize the results. Including zero modes, we find exactly $7B$ vibrational modes for the $B=1-4$ Skyrmions. We have included one mode (an $^1A_2$ mode for the $B=3$  Skyrmion) which lies slightly above $\omega = 1.5$. For the $B=5-8$ Skyrmions we find almost $7B$ modes, though never reach that number. The exact numbers can be found in Table \ref{NumOfVibs}. In theory there are infinitely many QNMs of higher and higher frequency and there is numerical evidence for this in the $B=1$ sector \cite{AHRW2}. Hence there is not necessarily a definite answer to how many modes a baryon number $B$ Skyrmion has. We believe Table \ref{NumOfVibs} has enough evidence to say that there are $7B$ low-lying modes and we conjecture that the missing modes lie somewhere beyond, but nearby, $\omega = 1.5$. Previous work has suggested that the deformation space has dimension $6B+1$. We can dismiss this due to the newly discovered modes, previously missed in Refs.~\cite{BBT2,BBT1}, for the $B=2$ and $4$ Skyrmions. An instanton-inspired approach suggests that $8B-1$ modes should be found \cite{AM}. We never find more than $7B$ modes and hence are doubtful there are $8B$ low-lying modes. Of course, since there are infinitely many modes, if one continues to search they will eventually find $8B$ modes. We conjecture that all of these will have significantly higher frequency than the modes we have found. Houghton showed that the instanton approach predicts a $^3T_1$ mode for the $B=3$ Skyrmion \cite{Hou}. Despite some effort searching, we have not found such a mode and instead have found a $^1A_2$ mode which is not predicted by this instanton approach. This suggests that the basic instanton approximation does not accurately model the local deformation space of Skyrmions.
	
	\begin{table}[!htp]
		\begin{center}
			\begin{tabular}{  l | c c c c c c c c c c }
				\hline
				$B$ & $1$ & $2$ & $3$ & $4$ & $5$ & $6$ & $7$ & $8_h$ & $8_t$ & $8_u$ \\ \hline \hline
				$N$ & $7$ & $14$ & $21^*$ & $28$ & $33$ & $41$ & $47$ & $53$ & $54$ & $54$ \\
				$7B - N$ & $0$ & $0$ & $0$ & $0$ & $2$ & $1$ & $2$ & $3$ & $2$ & $2$  \\ \hline \hline
			\end{tabular}
			\vskip 7pt
			\caption{The number $N$ of vibrational modes (including zero modes) found for each Skyrmion with $\omega < 1.5$. For the $B=3$ Skyrmion, a single mode at $\omega = 1.59$ is also included.}
			\label{NumOfVibs}
		\end{center}
	\end{table}
	
	At the frequency we search at, a $B=1$ Skyrmion has seven degrees of freedom: three translations, three rotations and a single breathing mode. Hence it is natural that a $B$ Skyrmion, being made from $B$ individual Skyrmions, has $7B$ degrees of freedom. If we increased our cut-off so that the $B=1$ could deform further (first through its dipole breathing mode of dimension three), the other deformation spaces may be enlarged proportionally.
	
	Generically the spectra split into four distinct parts, though the parts overlap in some highly symmetric cases. From low to high, the spectra begin with the zero-frequency modes: the translations, rotations and isorotations. Then there are the ``monopole modes". These can be described as the Skyrmion breaking into clusters where the individual clusters retain their orientation and size. There is a one-to-one mapping between these configuration spaces and the moduli space of monopoles, both of which can be described using rational maps \cite{HMS,LP}. There are then breathing modes where either different parts of the Skyrmion, or the entire Skyrmion, inflate and deflate. Finally, there are modes where different parts of the Skyrmion isorotate in different directions. These tend to have very high frequency.
	
	We list all of the vibrational modes discovered in Table \ref{modetable} alongside their corresponding frequencies and the irrep which they transform as. The different irreps for each Skyrmion are discussed and listed in Appendix \ref{app:A}. A description of each mode can be found in Appendix \ref{app:B} and animations of each mode can be viewed online at \website. Note that the normal modes of all the Skyrmions, apart from the $B=8_u$ Skyrmion, are real and positive. Hence, the Skyrmions are all locally stable apart from the $B=8_u$ Skyrmion, which decays into the $B=8_t$ Skyrmion through its single imaginary-valued mode.
	
	\begin{table}[!htp]
		\begin{center} 
			\begin{tabular}{@{}c@{\,}|@{\,}c@{\,}|llllllllllll}
				\hline
				B & \multicolumn{13}{c}{Vibrational modes in $d$ dimensional irrep $\mathcal{I}$ with frequency $\omega$}\\
				\hline\hline
				1 & $^d\mathcal{I}$ &
				$^1A_1$\\
				&$\omega$ &
				1.20\\
				\hline\hline
				2 & $^d\mathcal{I}$ &
				$^2E_{2g}$ & $^2E_{1u}$ & $^1A_{1g}$ & $^1A_{2u}$\\
				&$\omega$ &
				0.37      & 0.99      & 1.03 & 1.08 \\
				\hline\hline
				3 & $^d\mathcal{I}$ &
				$^3T_2$ & $^2E$ & $^3T_2$ &$^1A_1$  & $^2E$ & $^1A_2$\\
				&$\omega$ &
				0.43    & 0.56 & 0.91    & 0.94    & 1.01  & 1.59\\
				\hline\hline
				4 & $^d\mathcal{I}$ &
				$^2E_g$ & $^3T_{2g}$ & $^1A_{2u}$ & $^3T_{2u}$ & $^1A_{1g}$ & $^3T_{1u}$ & $^3T_{2g}$ & $^3T_{2u}$\\
				&$\omega$ &
				0.46    & 0.48      & 0.52      & 0.62      & 0.87      & 0.87      & 0.94      & 1.14\\
				\hline\hline
				5 & $^d\mathcal{I}$ &
				$^2E$ & $^1B_1$ & $^1A_1$ & $^1B_2$ & $^2E$ & $^1B_2$ & $^2E$ & $^1A_1$ & $^1B_1$ & $^1A_2$ & $^2E$ & $^1A_1$ \\
				&$\omega$ 
				& 0.10  & 0.23    & 0.42   & 0.43    & 0.45  & 0.49    & 0.51 & 0.57    & 0.61   & 0.77    & 0.80 & 0.81   \\
				&& $^1B_1$ & $^1B_2$ & $^1B_2$ & $^2E$ & $^1A_1$ & $^1A_2$ & $^1A_1$ \\
				&& 0.84   & 0.89    & 0.95  & 0.95   & 1.00    & 1.04    & 1.09  \\
				\hline\hline
				6 & $^d\mathcal{I}$ &
				$^1A_1$ & $^1B_1$ & $^2E_1$ & $^2E_3$ & $^2E_2$ & $^2E_2$ & $^1B_2$ & $^2E_1$ & $^2E_2$ & $^1A_1$ & $^2E_1$ & $^2E_3$ \\
				&$\omega$ 
				& 0.25  & 0.28    & 0.33      & 0.40       & 0.41    & 0.46   & 0.47    & 0.50       & 0.52    & 0.76    & 0.77 & 0.80      \\
				&& $^1B_2$ & $^2E_2$ & $^1A_1$ & $^2E_3$ & $^2E_1$ & $^2E_2$ & $^2E_3$\\
				&& 0.87    & 0.90    & 0.95      & 0.95    & 1.01      & 1.08      & 1.11  \\
				\hline\hline
				7 & $^d\mathcal{I}$ &
				$^5H_g$ & $^5H_u$ & $^2T_{2u}$ & $^4G_u$ & $^1A_{1g}$ & $^3T_{1u}$ & $^4G_g$ & $^5H_g$ & $^4G_u$ & $^4G_g$  \\
				&$\omega$ 
				& 0.32  & 0.38    & 0.47      & 0.54   & 0.73      & 0.76  & 0.84    & 0.91  & 0.94   & 1.01   \\
				\hline\hline
				$8_h$ & $^d\mathcal{I}$ &
				$^2E_3$ & $^2E_5$ & $^2E_4$ & $^2E_2$ & $^1B_1$ & $^2E_1$ & $^1A_1$ & $^1B_2$ & $^2E_2$ & $^2E_3$ & $^2E_2$ & $^2E_1$ \\
				&$\omega$ 
				& 0.13  & 0.21    & 0.24       & 0.28     & 0.33   & 0.40       & 0.45    & 0.45    & 0.45    & 0.49    & 0.50       & 0.57        \\
				&& $^1B_2$ & $^1A_1$ & $^2E_5$ & $^2E_1$ & $^2E_5$ & $^2E_4$ & $^1A_1$ & $^2E_4$ & $^2E_1$ & $^2E_2$ & $^2E_4$ & $^2E_5$ \\
				&& 0.69    & 0.70    & 0.75       & 0.76        & 0.87        & 0.88   & 0.89    & 0.89    & 0.90       & 0.96    & 1.01    & 1.02   \\
				&& $^2E_3$\\
				&& 1.03         \\
				\hline\hline
				$8_t$ & $^d\mathcal{I}$ &
				$^1A_{1u}$ & $^1A_{2g}$ & $^2E_u$ & $^1A_{1g}$ & $^1B_{1g}$ & $^1B_{1u}$ & $^2E_g$ & $^1A_{2u}$ & $^1A_{1g}$ & $^2E_u$ & $^1B_{2u}$ & $^1B_{1g}$ \\
				&$\omega$ 
				& 0.18    & 0.19      & 0.22    & 0.25      & 0.33      & 0.35      & 0.38   & 0.43      & 0.44       & 0.44    & 0.45 & 0.46      \\
				&& $^1B_{2g}$ & $^1B_{2g}$ & $^2E_g$ & $^1B_{1u}$ & $^2E_g$ & $^1B_{2u}$ & $^1A_{1g}$ & $^2E_u$ & $^1B_{1g}$ & $^1B_{2g}$ & $^1B_{1u}$ & $^1A_{2u}$ \\
				&& 0.47      & 0.48      & 0.48    & 0.52      & 0.59    & 0.62      & 0.72   & 0.72      & 0.81      & 0.83      & 0.84      & 0.85      \\
				&& $^2E_u$ & $^2E_g$ & $^1A_{1g}$ & $^2E_u$ & $^1A_{2u}$ & $^2E_{g}$ & $^1B_{1g}$ & $^1B_{1u}$ & $^1B_{2u}$ & $^2E_{u}$  \\
				&& 0.86    & 0.88   & 0.91      & 0.94    & 0.98      & 1.03      & 1.05      & 1.06 & 1.08     & 1.10  \\
				\hline\hline
				$8_u$ & $^d\mathcal{I}$ &
				$^1A_{2g}$ & $^1A_{1u}$ & $^2E_u$ & $^2E_g$ & $^1B_{2u}$ & $^1A_{1g}$ & $^1B_{2g}$ & $^1A_{1g}$ & $^1B_{1g}$ & $^1B_{1u}$ & $^2E_g$ & $^1A_{2u}$ \\
				&$\omega$ 
				& 0.12i    & 0.01  & 0.21    & 0.22    & 0.29     & 0.30       & 0.34      & 0.40      & 0.45       & 0.47   & 0.47     & 0.48      \\
				&& $^2E_u$ & $^1B_{1u}$ & $^1B_{2u}$ & $^1B_{2g}$ & $^1B_{1g}$ & $^1B_{2g}$ & $^2E_u$ & $^2E_g$ & $^1A_{1g}$ & $^1A_{2u}$ & $^2E_g$ & $^2E_u$ \\
				&& 0.49    & 0.51      & 0.52      & 0.53      & 0.57      & 0.57      & 0.68    & 0.70    & 0.78      & 0.85     & 0.86    & 0.86   \\
				&& $^1B_{1u}$ & $^1A_{1g}$ & $^2E_u$ & $^1B_{2u}$ & $^1B_{1g}$ & $^1A_{2u}$ & $^2E_g$ & $^1B_{2g}$ & $^1B_{2u}$ & $^2E_{u}$  \\
				&& 0.88      & 0.91       & 0.92   & 0.93      & 0.96      & 0.96      & 0.97    & 1.04      & 1.09      & 1.13 \\
				\hline\hline
			\end{tabular}
			\caption{A list of the frequencies and irreps of the numerically generated low lying normal and quasi-normal modes. The different irreps are explained in Appendix \ref{app:A} while a list of modes including descriptions can be found in Appendix \ref{app:B}.}
			\label{modetable}
		\end{center}
	\end{table}
	
	To describe excited states of nuclei one could promote the classical vibrations to quantum excitations in a harmonic approximation. Taking commonly used energy and length scales the excitations have energies much higher than those experimentally seen. This is related to another long-standing problem in the Skyrme model: that its classical binding energies appear too large. A modified model with low classical binding energies (such as those developed in Refs.~\cite{NS,GHS,Gud,Gud2,Gud3,Gud4,Gud5,ASW}) will also have lower vibrational frequencies and hence more realistic quantum states. In addition, and especially in the lightly bound models, the potential energy may be significantly anharmonic - a difficult feature to account for. However, a recently developed quantization method based on quantum graphs can deal with anharmonic potentials as well as other difficulties \cite{Raw}.
	
	Some particularly interesting vibrational modes are observed in the $B=8_t$ Skyrmion, the lowest-energy chain of two $B=4$ Skyrmions which are interpreted as alpha particles. The modes we refer to here are the lowest-frequency vibrational modes. The first mode is a $^1A_{1u}$ mode with $\omega=0.18$ which makes the two cubes rotate back and forth, out of phase, around their common axis. The next is a $^1A_{2g}$ mode with $\omega=0.19$. Here, the two cubes isorotate, out of phase, about the $\pi_1$ isospin axis. This is followed by a $^2E_u$ mode with $\omega=0.22$ which makes the two cubes wiggle into a bent-arm shape and then into the opposite bent-arm shape. The next mode is a $^1A_{1g}$ mode with $\omega=0.25$ and here the two cubes move away from each other and back again. Finally, the next mode is a $^1B_{1g}$ mode with $\omega=0.33$ which makes the two cubes isorotate around the $\pi_3$ axis, out of phase. These first five vibrational modes excite the zero modes of the individual alpha particles: their relative motion and (iso)rotations. The sixth vibrational mode includes vibrational modes of the alpha particles themselves as the frequency is now high enough that the structure of the alpha particle can be seen. This makes a clear physical picture of vibrations of the cluster-like Skyrmions. Their lowest-lying modes are just movements and (iso)rotations of the individual clusters while their higher-frequency modes include vibrations of the alpha particles themselves. 
	
\begin{figure}[!ht]
		\begin{center}
			\includegraphics[width=\linewidth]{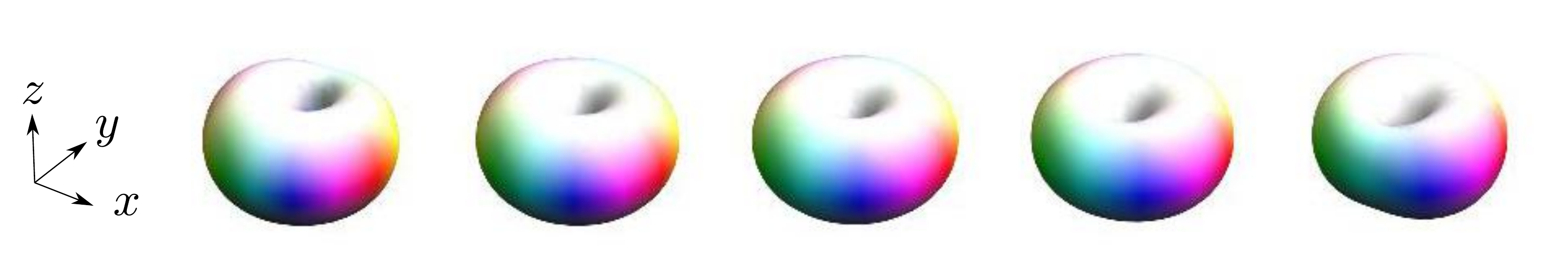}
			\caption{The $\omega=0.99$ mode of the $B=2$ Skyrmion which transforms as $^2E_{1u}$. It can be interpreted as either a dipole breather or a non-global isorotation.}
			\label{fig:B2Eu99}
		\end{center} 
	\end{figure}

	A high frequency mode which is difficult to interpret is displayed in Figure \ref{fig:B2Eu99}. It is a mode of the $B=2$ Skyrmion lying at $\omega=0.99$ which transforms as the $^2E_{1u}$ irrep. Looking only at its energy density it appears that one half of the Skyrmion inflates while the other half deflates. This was the interpretation given in Ref.~\cite{BBT2}. However, examining the pion field structure we can see that the red color (where $\hat{\pi}_1$ = 1) moves from one side to another, as does the the red on the other side of the Skyrmion. If one imagines the $B=2$ Skyrmion as two $B=1$ Skyrmions slightly separated along the $x$-axis, this pion field motion is consistent with the individual Skyrmions isorotating about the $\pi_3$-axis, out of phase. This isorotation distorts the energy density. This mode demonstrates the difficulty of interpreting modes - is this a breather mode or an out-of-phase isorotation? This dual interpretation could be a reflection of the fact that many solitons cannot be easily described in terms of their individual components.
	
	\section{Conclusions and Further Work} \label{sec:4}
	
	We have generated and classified the low-lying normal and quasi-normal modes of the $B=1-8$ Skyrmions. Based on these calculations we conjecture that there are $7B$ such modes for each Skyrmion. This conjecture would be confirmed if one could find the missing modes from Table \ref{NumOfVibs}. Unfortunately, as $\omega$ is increased the density of scattering modes also increases, making the identification of QNMs significantly more difficult. Ideally a different method, specifically designed to find QNMs, should be used.
	
	To fully understand the Skyrmion configuration space and do calculations on it, an approximation of the space is needed. We now know to look for a $7B$-dimensional approximation. It may be possible to restrict the (holonomy of the) instanton moduli space, which is $(8B-1)$-dimensional, to describe such a space. Or an entirely new approach is required. The data in Table \ref{modetable} provides a stringent test of any proposed approximation. It should be able to reproduce the irrep decomposition of the modes for all of the Skyrmions listed. 
	
	Each mode we have generated may be important for specific nuclear processes and properties. For instance, most of the low-lying Oxygen-$16$ energy spectrum can be understood in terms of a single normal mode once its full configuration space is studied and understood \cite{HKM}. Our hope is that the modes found here provide a useful foundation for similar calculations. Each mode listed in Appendix \ref{app:B} is the first hint of the structure of a full nonlinear deformation space. Many of these will have interesting geometric properties, as has been the case in those studied so far \cite{Hal,HKM}. The modes may also provide a path towards saddle point solutions in each topological sector. Understanding these structures and solutions is a crucial step towards an accurate description of nuclei within the Skyrme model.
	
	\subsection*{Acknowledgments}
	
	We thank Nick Manton for discussions.
	S.~B.~G.~is supported by the Ministry of Education, Culture, Sports,
	Science (MEXT)-Supported Program for the Strategic Research Foundation
	at Private Universities ``Topological Science'' (Grant No.~S1511006) and
	by a Grant-in-Aid for Scientific Research on Innovative Areas
	``Topological Materials Science'' (KAKENHI Grant No.~15H05855) from
	MEXT, Japan. C.~H.~is supported by The Leverhulme Trust as an Early Careers Fellow.  C.~H. would like to thank Muneto Nitta for hospitality during his visit to Keio University, where much of this work was completed. The numerical calculations for this work were carried out using the TSC computing 
	cluster of the ``Topological Science'' project at Keio University.

	\appendix
	
	\section{Symmetries} \label{app:A}
	
	Each Skyrmion has a symmetry group generated by combined rotation-isorotation operators. The modes are then classified by the irreducible representations (irreps) of the groups. The symmetry operators commute with the evolution operator \eqref{epeom} and so if an initial perturbation transforms as a certain irrep, the evolved perturbation will as well. Modes which transform as different irreps are automatically orthogonal. Hence we can study each irrep separately. Doing so means that we study deformation spaces of lower dimension. These are easier to interpret as the modes are generally more sparsely spaced in frequency. This is advantageous as our method converges as $\exp(-\tau|\omega_i^2 - \omega_{i+1}^2|)$ for the $i^\text{th}$ mode. In addition, the QNMs are much easier to identify. Looking again at Figure \ref{fig:quasiev}, the existence of the $^1B_{1g}$ QNM is rather clear. If the two spectra were combined the identification would be less obvious.
	
	For multi-dimensional irreps, we look for a specific realization of the irrep. The choice of realization depends on the situation at hand. If a two-dimensional irrep transforms like the Cartesian coordinates $(x,y)$ we will sometimes choose to look for elements which transform like $x$ and sometimes for elements which transform as $x-y$.
	
	In practice our numerical scheme takes place in a rectangular box. Hence the full symmetry group is not preserved on the lattice -- only the subgroup compatible with the lattice symmetries is realized numerically. The full group is preserved for the $B=3, 4, 5, 8_u \text{ and } 8_t$ Skyrmions. For other Skyrmions, such as the $B=6$, the full group is broken. In this case, the $D_{4d}$ group is reduced to the $D_4$ subgroup. We must then make a correspondence between the irreps of each group. For instance, the $^1A_1$ irrep of the $D_4$ group is compatible with both the $^1A_1$ and $^1B_1$ irreps of the $D_{4d}$ group. Hence we cannot search for these irreps separately and must instead search for them together by studying the modes which transform as the $^1A_1$ irrep of the $D_4$ group. Once found, these can be further classified by applying approximate symmetry operators to the generated modes.
	
	Details of the symmetry groups and their irreps can be found in Table \ref{tab:symm}. We have chosen to adopt the notation of Ref.~\cite{Cot} which is widely used online. The character tables of the groups can be found in Ref.~\cite{Cot} or at \href{http://symmetry.jacobs-university.de}{http://symmetry.jacobs-university.de}.
	
	\begin{table}[!htp]
		\begin{center} 
			\begin{tabular}{ l | l | l | c } \hline 
				$B$ & $\mathcal{G}$ & $\mathcal{H}$  & $^d\mathcal{I}$ \\ \hline \hline
				2 & $D_{\infty h}$ & $D_{4h}$ & $^1A_{1g},\, ^1A_{2g},\, ^2E_{1g},\, ^2E_{2g},\, \ldots,\, ^1A_{1u},\, ^1A_{2u},\, ^2E_{1u},\, ^2E_{2u},\, \ldots$ \\
				3 & $T_d$ & $T_d$ & $^1A_1,\, ^1A_2,\, ^2E,\, ^3T_1,\, ^3T_2$\\
				4 & $O_h$ & $O_h$ &  $^1A_{1g},\, ^1A_{2g},\, ^2E_g,\, ^3T_{1g},\, ^3T_{2g},\,^1A_{1u},\, ^1A_{2u},\, ^2E_u,\, ^3T_{1u},\, ^3T_{2u}$\\
				5 & $D_{2d}$ &  $D_{2d}$ & $^1A_1,\, ^1A_2,\, ^1B_1,\, ^1B_2,\, ^2E$ \\
				6 & $D_{4h}$ & $D_4$ & $^1A_1,\, ^1A_2,\, ^1B_1,\, ^1B_2,\, ^2E_1,\, ^2E_2,\, ^2E_3$ \\
				7 & $I_h$ & $T_h$ & $^1A_{g},\, ^3T_{1g},\, ^3T_{2g},\, ^4G_{g},\, ^5H_g,\, ^1A_{u},\, ^3T_{1u},\, ^3T_{2u},\, ^4G_{u},\, ^5H_u$ \\
				$8_h$ & $D_{6d}$ & $D_{2d}$ & $^1A_1,\, ^1A_2,\, ^1B_1,\, ^1B_2,\, ^2E_1,\, ^2E_2,\, ^2E_3,\, ^2E_4,\, ^2E_5$  \\
				$8_t$, $8_u$  & $D_{4h}$ & $D_{4h}$ & $^1A_{1g},\, ^1A_{2g},\, ^1B_{1g},\, ^2B_{2g},\, ^2E_{g},\, ^1A_{1u},\, ^1A_{2u},\, ^1B_{1u},\, ^2B_{2u},\, ^2E_{u}$\\ \hline \hline
				
			\end{tabular}
			\caption{The symmetry groups, $\mathcal{G}$ of each Skyrmion, their lattice-compatible subgroup $\mathcal{H}$ and a list of $\mathcal{G}$'s irreps $\mathcal{I}$ and their dimension $d$.}
			\label{tab:symm}
		\end{center}
	\end{table}

	\section{The vibrational modes} \label{app:B}
	
	Here, we list the normal modes and QNMs of the $B=1-8$ Skyrmions. Their frequency, dimension, irrep and a description can be found in Table \ref{tab:modes}. When describing modes which fall into a multi-dimensional irrep, we usually only describe one possible realization of the mode.
	
	\begin{tabularx}{\textwidth}{ c  c  c  X   } 
		\hline
		$B$&$\omega$ & $^d$Irrep & Description \\ \hline \hline

		$1$ & 1.24 & $^1A_{1g}$ & The breather. \\ \hline \hline

		$2$ & 0.37 & $^2E_{2g}$  &  The torus splits into two individual Skyrmions.\\
		& 0.99 & $^2E_{1u}$  &  Two halves of the torus isorotate in different directions, around the $\pi_3$ axis. This may also be interpreted as two individual Skyrmions breathing out of phase. \\
		& 1.03 & $^1A_{1g}$  &  The breathing mode.\\
		& 1.08 & $^1A_{2u}$  &  The equator of the color wheel moves up and down.\\ \hline \hline

		3  & 0.43 & $^3T_2$  &  A vertex of the tetrahedron moves away from its opposing face. Asymptotically, these become separate $B=1$ and $B=2$ clusters.\\
		& 0.56 & $^2E$  &  Two opposite edges of the tetrahedron pull away from each other, deforming it into a pretzel-like shape. This space contains the well-known twisted-line scattering \cite{HS}.\\
		& 0.91 & $^3T_2$  &  One vertex grows while the other three shrink.\\
		& 0.94 & $^1A_1$  &  The breathing mode.\\
		& 1.01 & $^2E$  &  Two vertices isorotate out of phase with the other two.\\
		& 1.59 & $^1A_2$  &  Each vertex isorotates about the color on its tip.\\ \hline \hline
		
		4  & 0.46 & $^2E_g$  &  Two opposite faces pull away from each other to form two $B=2$ tori. In the other direction, four edges pull away to become four $B=1$ Skyrmions.\\
		& 0.48 & $^3T_{2g}$  &  An opposing pair of square-symmetric faces deform to become rhombus shaped.\\
		& 0.52 & $^1A_{2u}$  &   Four vertices of the cube pull away, retaining tetrahedral symmetry. These then come in again and the other four vertices pull away to form the dual tetrahedron.\\
		& 0.62 & $^3T_{2u}$  &  Two opposite edges from the same face pull away from the origin. On the opposite face, the perpendicular edges also pull away.\\
		& 0.87 & $^1A_{1g}$  &  The breathing mode.\\
		& 0.87 & $^3T_{1u}$  &  One face inflates while the opposite face deflates.\\
		& 0.94 & $^3T_{2g}$  &  Two opposite faces isorotate in opposite directions. \\
		& 1.14 & $^3T_{2u}$  &  Four vertices (all lying in a plane that also goes through the origin) isorotate clockwise around the $\pi_1$ axis. The other four isorotate anti-clockwise.\\
		\hline \hline

		5  & 0.10 & $^2E$  &  The top edge shears away from the rest of the Skyrmion, which moves to preserve the center of mass. The bottom edge remains stationary.\\
		& 0.23 & $^1B_1$  &  The top and bottom edges twist in opposite directions.\\
		& 0.42 & $^1A_1$  &  The entire Skyrmion elongates. Asymptotically, it becomes a chain of five individual Skyrmions.\\
		& 0.43 & $^1B_2$  &  The Skyrmion splits in half, four of its holes lying in the incision plane.\\
		& 0.45 &$^2E$  &  A $B=2$ torus emerges from the top of the Skyrmion. Asymptotically, the Skyrmion splits into $B=2$ and $B=3$ clusters.\\
		& 0.49 & $^1B_2$  &  The top edge detaches from the rest, leaving a B=4 core. In the other direction, the bottom edge detaches instead.\\
		& 0.51 & $^2E$  &  A $B=$1 Skyrmion detaches from the equatorial torus.\\
		& 0.57 & $^1A_1$  &   The Skyrmion elongates while also pushing outwards. The asymptotic configuration looks like a riding saddle with $D_{2d}$ symmetry.\\
		& 0.61 &$^1B_1$  &  The equator, interpreted as a $B=2$ Skyrmion, performs $1+1$ 90$^\circ$  scattering.\\
		& 0.77 & $^1A_2$  &  The top and bottom halves of the Skyrmion twist in different directions.\\
		& 0.8 & $^2E$  &  One of the four long faces (containing two holes) inflates while the opposite one deflates. The remaining two remain unchanged.\\
		& 0.81 & $^1A_1$  &  The top and bottom edges break away from the equator, which deforms into a square-like object.\\
		& 0.84 & $^1B_1$  &  The top half isorotates clockwise around the $\pi_3$ axis while the bottom half isorotates in the other direction.\\
		& 0.89 & $^1B_2$  &  The top half inflates while the bottom half deflates.\\
		& 0.95 & $^1B_2$  &  Similar to the 0.43 mode but physically due to breathing.\\
		& 0.95 & $^2E$  &  Similar to the 0.45 mode but physically due to breathing.\\
		& 1.00 & $^1A_1$  &  The breathing mode.\\
		& 1.04 & $^1A_2$  &  The top and bottom half isorotate in the same direction, but the equator remains unchanged. This is similar to the 0.84 mode but in phase.\\
		& 1.09 & $^1A_1$  &  Similar to the 0.57 mode but physically due to breathing.\\ \hline \hline

		6 & 0.25 & $^1A_1$  &  The outer tori move away from the central torus.\\
		& 0.28 & $^1B_1$  &  The outer tori rotate in opposite directions.\\
		& 0.33 & $^2E_1$  &  The outer tori rotate, deforming the Skyrmion like a hinge.\\
		& 0.40 & $^2E_3$ &  The outer tori shear, out of phase.\\
		& 0.41 & $^2E_2$  &  The outer tori split into two $B=1$ Skyrmions in the plane perpendicular to their common symmetry axis. The bottom mirrors the motion of the top, but rotated by 45$^\circ$. In both cases, the $B=1$'s move towards a vertex of the core.\\
		& 0.46 & $^2E_2$  &  A similar motion as the 0.41 mode, but the $B=1$'s move towards a hole of the core.\\
		& 0.47 & $^1B_2$  &  One of the outer tori pull away, leaving a $B=4$ core.\\
		& 0.50 & $^2E_1$  &   The center torus pulls away from the Skyrmion while the outer tori deform to maintain the center of mass.\\
		& 0.52 & $^2E_2$  &  Viewing the $B=6$ Skyrmion as two overlapping $B=4$ cubes, the cubes perform the $B=4$, 0.52 mode out of phase.\\
		& 0.76 & $^1A_1$  &  The breathing mode.\\
		& 0.77 & $^2E_1$ &  Two of the eight central holes grow while another two shrink.\\
		& 0.80 & $^2E_3$  &  Viewing the $B=6$ Skyrmion as two overlapping $B=4$ cubes, the cubes perform the $B=4$, 0.62 mode out of phase.\\
		& 0.87 & $^1B_2$  &  One half of the Skyrmion deflates as the other inflates. The central torus oscillates between the two outer tori, yielding a Newton's cradle motion of tori.
		\\
		& 0.90 & $^2E_2$  &  Similar to the 0.46 mode but physically due to breathing.\\
		& 0.95 & $^1A_1$ &  The Skyrmion elongates then flattens, maintaining $D_{4h}$ symmetry.\\
		& 0.95 & $^2E_3$  &  One face inflates while the diagonally opposite face deflates.\\
		& 1.01 & $^2E_1$  &  Viewing the $B=6$ Skyrmion as two overlapping $B=4$ cubes, the cubes perform the $B=4$, 0.94 mode out of phase.\\
		& 1.08 & $^2E_2$  &  Similar to the 0.41 mode but physically due to isorotation.\\
		& 1.11 & $^2E_3$ &  Viewing the $B=6$ Skyrmion as two overlapping $B=4$ cubes, the cubes perform the $B=4$, 1.14 mode in phase.\\ \hline \hline

		7  & 0.32 & $^5H_g$  &  Two opposite faces pull away from the center of the Skyrmion.\\
		& 0.38 & $^5H_u$ &  The energy density concentrates around an edge. Asymptotically this becomes the edge of a $B=4$ Skyrmion. The opposite side of the Skyrmion becomes a $B=3$ torus.\\
		& 0.47 & $^3T_{2u}$  &  Two nearby faces pull away from the center. The remaining energy density forms a long hat on the opposite side of the Skyrmion. Asymptotically, it is pulling out 3 tori and leaving a 1-Skyrmion behind. \\
		& 0.54 & $^4G_u$  &  The dodecahedron contains five cubes. Here, a single cube is deformed while retaining tetrahedral symmetry.\\
		& 0.73 & $^1A_g$  &  The breathing mode.\\
		& 0.76 & $^3T_{1u}$  &  A dipole breathing mode.\\
		& 0.84 & $^4G_g$  &  This is the mode described by Singer and Sutcliffe \cite{SS}. Asymptotically, six individual Skyrmions travel along the Cartesian axes towards one at the origin. They form the dodecahedron then become a cube. Our deformation is not large enough to reach the cubic structure. The dodecahedron contains five cubes and so there are naively five of these modes. However, they are linearly dependent as the sum of all five modes is trivial.\\
		& 0.91 & $^5H_g$  &  Similar to the 0.32 mode but physically due to breathing.\\
		& 0.94 & $^4G_u$  &  A non-trivial out-of-phase isorotation which retains the tetrahedral symmetry of one of the dodecahedron's cubes.\\
		& 1.01 & $^4G_g$  &  Another non-trivial out-of-phase isorotation which retains the tetrahedral symmetry of one of the dodecahedron's cubes, but in a different way than the 0.94 mode.\\ \hline \hline
		
		$8_h$	  & 0.13 & $^2E_3$  &  The top and bottom tori slide back and forth, out of phase, yielding a shear mode.\\
		& 0.21 & $^2E_5$ &  The top and bottom tori rotate around an axis perpendicular to their symmetry axis, out of phase.\\
		& 0.24 & $^2E_4$  &  The Skyrmion begins to split in half, along the x-axis and then the y-axis.\\
		& 0.28 & $^2E_2$  &  Two opposite vertices of the equatorial hexagon move upwards while the edges which lie between them move downwards.\\
		& 0.33 & $^1B_1$  &  The top and bottom tori rotate, out of phase.\\
		& 0.40 & $^2E_1$ &  A single Skyrmion lying on a vertex of the equatorial hexagon pulls away from the center. The remaining Skyrmion deforms to compensate for this motion. \\
		& 0.45 & $^1A_1$  &  The top and bottom tori move away from the center at the same time.\\
		& 0.45 & $^1B_2$  &  The top and bottom tori move away from, and then towards the center, out of phase.\\
		& 0.45 & $^2E_2$  &  The top and bottom tori split into individual Skyrmions, out of phase.\\
		& 0.49 & $^2E_3$  &  The tori move from side to side in phase, while the equator moves in the opposite direction to keep the center of mass constant.\\
		& 0.5 & $^2E_2$  &  The top and bottom tori split into individual Skyrmions, in phase. The equator deforms significantly to accommodate the splitting.\\
		& 0.57 & $^2E_1$  &  Two sides of the hexagonal equator breathe, out of phase.\\
		& 0.69 & $^1B_2$  &  The top and bottom tori breathe, out of phase.\\
		& 0.70 & $^1A_1$  &  The entire Skyrmion breathes.\\
		& 0.75 & $^2E_5$  &  Two opposite quarters of the Skyrmion inflate, pushing the torus closest to them away. \\
		& 0.76 & $^2E_1$  &  One half of the Skyrmion inflates, pushing both tori away from the center. \\
		& 0.87 & $^2E_5$  &  Two holes which lie on an axis stay stationary. The neighboring holes move towards one and away from the other.\\
		& 0.88 & $^2E_4$  &  The top and bottom tori split into individual Skyrmions, out of phase. Unlike the 0.50 mode, the equator does not deform. \\
		& 0.89 & $^1A_1$  &  The top and bottom tori breathe, in phase.\\
		& 0.89 & $^2E_4$  & The equator inflates along one axis and deflates along the perpendicular one.\\
		& 0.90 & $^2E_1$  &  Similar to the 0.40 mode but physically due to isorotation.\\
		& 0.96 & $^2E_2$  &  Two neighboring faces move towards, then away, from one another.\\
		& 1.01 & $^2E_4$  &  Similar to the 0.24 mode but physically due to isorotation.\\
		& 1.02 & $^2E_5$  &  Similar to the 0.21 mode but physically due to isorotation.\\
		& 1.03 & $^2E_3$  &  A tri-axially symmetric breathing motion.\\ \hline \hline

		$8_t$  & 0.18 & $^1A_{1u}$  &  The cubes rotate around their common $C_4$ symmetry axis, out of phase.\\
		& 0.19 & $^1A_{2g}$  &  The cubes isorotate around the $\pi_1$ iso-axis, out of phase.\\
		& 0.22 & $^2E_u$  &  The cubes rotate towards each other forming a bent-arm.\\
		& 0.25 & $^1A_{1g}$  &  The cubes move away from each other.\\
		& 0.33 & $^1B_{1g}$  &  The cubes isorotate around the $\pi_3$ iso-axis, out of phase.\\
		& 0.35 & $^1B_{1u}$  &  Each cube vibrates like the $B=4$, 0.48 mode out of phase.\\
		& 0.38 & $^2E_g$  &  Similar to the 0.22 mode, but the rotation is in phase yielding a shear mode of the two cubes.\\
		& 0.43 & $^1A_{2u}$  &  The cubes each vibrate like the $B=4$, 0.46 mode out of phase. The outgoing tori lie in the plane perpendicular to the common $C_4$ symmetry axis.\\
		& 0.44 & $^1A_{1g}$  &  The central cube vibrates like the $B=4$, 0.46 mode. \\
		& 0.44 & $^2E_u$  &  Each cube vibrates like the $B=4$, 0.48 mode out of phase.\\
		& 0.45 & $^1B_{2u}$  &  Each cube vibrates like the $B=4$, 0.46 mode out of phase. \\
		& 0.46 & $^1B_{1g}$  &  Each cube vibrates like the $B=4$, 0.46 mode in phase.\\
		& 0.47 & $^1B_{2g}$  &  The outer tori vibrate like the $B=2$, 0.37 mode in phase.\\
		& 0.48 & $^1B_{2g}$  &  The central cube vibrates like the $B=4$, 0.48 mode.\\
		& 0.48 & $^2E_g$  &  Each cube vibrates like the $B=4$, 0.48 mode, in phase.\\
		& 0.52 & $^1B_{1u}$  &  Each cube vibrates like the $B=4$, 0.52 mode, in phase.\\
		& 0.59 & $^2E_g$  &  Each cube vibrates like the $B=4$, 0.62 mode out of phase realized in such a way that the central cube's deformation is small.\\
		& 0.62 & $^1B_{2u}$  &  Each cube vibrates like the $B=4$, 0.62 mode in phase realized in such a way that the central cube significantly deforms.\\
		& 0.72 & $^1A_{1g}$  &  The center cube breathes.\\
		& 0.72 & $^2E_u$  &  Each cube vibrates like the $B=4$, 0.94 mode, realized such that two edges of the central cube pull away from each other.\\
		& 0.81 & $^1B_{1g}$  &  The central cube vibrates like the $B=4$, 0.87 mode while the top and bottom tori vibrate like the $B=2$, 0.37 mode. All motion is due to breathing. \\
		& 0.83 & $^1B_{2g}$  &  Each cube vibrates like the $B=4$, 0.94 mode in phase.\\
		& 0.84 & $^1B_{1u}$  &  The central cube vibrates like the $B=4$, 0.87 mode.\\
		& 0.85 & $^1A_{2u}$  &  One cube inflates while the other deflates. The central cube moves back and forth between the two tori.\\
		& 0.86 & $^2E_u$  &  Two neighboring faces inflate while the opposite faces deflate. The faces in between move their positions to compensate.\\
		& 0.88 & $^2E_g$  &  Each cube vibrates like the $B=4$, 0.87 mode out of phase.\\
		& 0.91 & $^1A_{1g}$  &  Both cubes breathe in phase.\\
		& 0.94 & $^2E_u$  &  Similar to the 0.22 mode but physically due to isorotation.\\
		& 0.98 & $^1A_{2u}$  &  The top and bottom tori inflate and deflate out of phase.\\
		& 1.03 & $^2E_g$  &  Each cube vibrates like the $B=4$, 0.94 mode out of phase.\\
		& 1.05 & $^1B_{1g}$  &  Similar to the 0.46 mode but physically due to a breathing motion.\\
		& 1.06 & $^1B_{1u}$  &  Similar to the 0.35 mode but physically due to a breathing motion.\\
		& 1.08 & $^1B_{2u}$  &  Similar to the 0.45 mode but physically due to a breathing motion.\\
		& 1.10 & $^2E_u$  &  The two cubes vibrate like the $B=4$, 1.14 mode in phase.\\ \hline \hline 
		
		$8_u$ & 0.12i & $^1A_{2g}$  &  Each cube isorotates around the $\pi_1$ isospin axis, in opposite directions. This mode connects the $B=8_u$ Skyrmion to the lower energy $B=8_t$ Skyrmion, hence its imaginary frequency.\\
		& 0.01 & $^1A_{1u}$  &  The cubes rotate around their common $C_4$ symmetry axis, out of phase.\\
		& 0.21 & $^2E_u$  &  The cubes rotate towards each other, making a bent-arm shape.\\
		& 0.22 & $^2E_g$  &  The cubes rotate away from each other, yielding a shear mode.\\
		& 0.29 & $^1B_{2u}$  &  The cubes isorotate about the white/black axis out of phase.\\
		& 0.30 & $^1A_{1g}$  &  The cubes move away from each other.\\
		& 0.34 & $^1B_{2g}$  &  The outer tori vibrate like the $B=2$, 0.37 mode in phase.\\
		& 0.40 & $^1A_{1g}$  &   Each cube vibrates like the $B=4$, 0.46 mode. The individual tori come out along the $C_4$ symmetry axis.\\
		& 0.45 & $^1B_{1g}$  &  Each cube vibrates like the $B=4$, 0.46 mode in phase. The tori come out along the axes perpendicular to the $C_4$ symmetry axis.\\
		& 0.47 & $^1B_{1u}$  &  The outer tori vibrate like the $B=2$, 0.37 mode out of phase.\\
		& 0.47 & $^2E_g$  &  Each cube vibrates like the $B=4$, 0.48 mode in phase.\\
		& 0.48 & $^1A_{2u}$  &  The two cubes vibrate like the $B=4$, 0.46 mode out of phase.\\
		& 0.49 & $^2E_u$  &  Each cube vibrates like the $B=4$, 0.48 mode out of phase.\\
		& 0.51 & $^1B_{1u}$  &  The central cube vibrates like the $B=4$, 0.52 mode.\\
		& 0.52 & $^1B_{2u}$  &  Each cube vibrates like the $B=4$, 0.46 mode out of phase.\\
		& 0.53 & $^1B_{2g}$  &  Each cube vibrates like the $B=4$, 0.52 mode out of phase.\\
		& 0.57 & $^1B_{1g}$  &  Each cube vibrates like the $B=4$, 0.62 mode such that one cube is a mirror image of the other.\\
		& 0.57 & $^1B_{2g}$  &  Each cube vibrates like the $B=4$, 0.48 mode in phase.\\
		& 0.68 & $^2E_u$  &  Each cube vibrates like the $B=4$, 0.62 mode such that one cube is a mirror image of the other.\\
		& 0.70 & $^2E_g$  &  Each cube vibrates like the $B=4$, 0.62 mode out of phase.\\
		& 0.78 & $^1A_{1g}$  &  The central cube breathes.\\
		& 0.85 & $^1A_{2u}$  &  Similar to the 0.40 mode but physically due to a breathing motion.\\
		& 0.86 & $^2E_g$  &  Each cube vibrates like the $B=4$, 0.87 $^3T_{1u}$ mode out of phase.\\
		& 0.86 & $^2E_u$  &  The central cube vibrates like the $B=4$, 0.87 $^3T_{1u}$ mode.\\
		& 0.88 & $^1B_{1u}$  &  Each cube vibrates like the $B=4$, 0.94 mode out of phase.\\
		& 0.91 & $^1A_{1g}$  &  Each cube breathes, in phase.\\
		& 0.92 & $^2E_u$  &  Similar to the 0.68 mode but physically due to an isorotation.\\
		& 0.93 & $^1B_{2u}$  &  The cubes vibrate like the $B=4$, 0.94 mode in phase.\\
		& 0.96 & $^1B_{1g}$  &  Similar to the 0.45 mode but physically due to a breathing motion.\\
		& 0.96 & $^1A_{2u}$  &  Similar to the 0.48 mode but physically due to a breathing motion.\\
		& 0.97 & $^2E_g$  &  Each cube vibrates like the $B=4$, 0.94 mode in phase.\\
		& 1.04 & $^1B_{2g}$  &  Similar to the 0.53 mode but physically due to a breathing motion.\\
		& 1.09 & $^1B_{2u}$  &  The central cube vibrates like the $B=4$, 0.94 mode.\\
		& 1.13 & $^2E_u$  &  Each cube vibrates like the $B=4$, 1.14 mode in phase.\\
		
		\hline \hline

		\caption{The modes of the $B=1-8$ Skyrmions.}
		\label{tab:modes}
	\end{tabularx}

\end{document}